\documentclass[apjl]{emulateapj}
\usepackage{natbib}
\usepackage{hyperref}
\hypersetup{
    colorlinks=true,
    linkcolor=blue,
    citecolor=blue,
    filecolor=blue,
    urlcolor=blue
}
 \providecommand{\adsurl}[1]{\href{#1}{ADS}}

\usepackage{amsmath,amssymb}
\usepackage{mathptmx}
\usepackage{mathtools}
\DeclareMathAlphabet{\pazocal}{OMS}{zplm}{m}{n}

\usepackage{graphicx}

\newcommand\ho{\ifmmode {\rm HI} \else H{\small I} \fi}
\newcommand\hh{\ifmmode {\rm H_2} \else H$_2$ \fi}
\newcommand\hii{\ifmmode {\rm HII} \else H{\small II} \fi}
\def\no{\ifmmode {N_{\rm HI}} \else $N_{\rm HI}$ \fi}
\def\nt{\ifmmode {N_{\rm H_2}} \else $N_{\rm HI}$ \fi}
\def\sho{\ifmmode {\Sigma_{\rm HI}} \else $\Sigma_{\rm HI}$ \fi}
\def\sob{\ifmmode {\Sigma_{\rm HI,obs}} \else $\Sigma_{\rm HI,obs}$ \fi}
\def\msun{\ifmmode {\rm M_{\odot}}\else $\rm M_{\odot}$\fi}
\def\mpc{\ifmmode {\rm M_{\odot} \ pc^{-2}} \else $\rm M_{\odot} \ pc^{-2}$ \fi}
\def\tra{\ifmmode  \text{HI-to-H}_2\else H{\small I}-to-H$_2$ \fi}
\def\aG{\ifmmode {\alpha G}\else $\alpha G$ \fi}
\def\iuv{\ifmmode {I_{\rm UV}}\else $I_{\rm UV}$ \fi}

\def\sg{\ifmmode \sigma_g \else $\sigma_g$ \fi}
\def\st{\ifmmode {\widetilde \sigma}_g \else ${\widetilde \sigma}_g$ \fi}
\newcommand\hd{\ifmmode \textrm{HI-dust} \else H{\small I}-dust \fi}
\newcommand\Nh{\ifmmode \pazocal{N} \else $\pazocal{N}$ \fi}

\defcitealias{Sternberg2014}{S14}
\defcitealias{Bialy2016}{BS16}
\defcitealias{Krumholz2009}{K09}
\defcitealias{Bihr2015}{B15}
\defcitealias{Motte2014}{M14}

\begin{document}

\title{\tra Transition Layers in the Star-Forming Region W43}
 \author{Shmuel Bialy$^\star$\altaffilmark{1}, Simon Bihr\altaffilmark{2}, Henrik Beuther\altaffilmark{2}, Thomas Henning\altaffilmark{2}, and Amiel Sternberg\altaffilmark{1}}
 \altaffiltext{1}
 {Raymond and Beverly Sackler School of Physics \& Astronomy,
 Tel Aviv University, Ramat Aviv 69978, Israel}
 \altaffiltext{2}
 {Max Planck Institute for Astronomy, K{\"o}nigstuhl 17, 69117 Heidelberg, Germany}
 \email{$^\star$shmuelbi@mail.tau.ac.il}

\begin{abstract}
The process of atomic-to-molecular ($\tra$) gas conversion is fundamental for molecular-cloud formation and star formation.
21 cm observations of the star-forming region W43 revealed extremely high \ho column densities, of 120-180 $\mpc$, a factor of 10-20 larger than predicted by \tra transition theories.
We analyze the observed \ho with an \tra transition theoretical model, and show that the theory-observation discrepancy cannot be explained by the intense radiation in W43, nor by variations of the assumed volume density or H$_2$ formation-rate coefficient.
We show that the large observed \ho columns are naturally explained by several ($9-22$) \tra transition layers,  superimposed along the sightlines of W43. 
We discuss other possible interpretations such as a non-steady-state scenario, and inefficient dust absorption. 
The case of W43 suggests that \ho thresholds reported in extra-galactic observations are probably not associated with a single \tra transition, but are rather a result of several transition layers (clouds) along the sightlines, beam-diluted with diffuse inter-cloud gas.
\end{abstract}

\section{Introduction}
% Giant molecular clouds (GMCs), shielded of the destructive UV radiation are the stellar nurseries of our Galaxy.
The transition of interstellar gas from atomic ($\ho$) to molecular (H$_2$) form is of fundamental importance for the process of star formation, and has been studied via analytic modeling
\citep[e.g.,][]{Federman1979, Sternberg1989, Goldsmith2007, Krumholz2008, McKee2010, Sternberg2014, Liszt2015, Bialy2016}, hydrodynamical simulations \citep[e.g.,][]{Robertson2008, Gnedin2009, Lagos2015, Valdivia2016} and observations \citep[e.g.,][]{Savage1977, Reach1994, Gillmon2006, Lee2012, Noterdaeme2016}.

\citet[][hereafter \citetalias{Motte2014}]{Motte2014} and  \citet[][hereafter \citetalias{Bihr2015}]{Bihr2015}  studied the \ho and \hh gas in the W43 star-forming complex.
W43 ($l = 29.2-31.5^{\circ}$, $|b| \leq 1^{\circ}$) is a very active star-formation region, containing many molecular clouds as well as atomic gas, with a total atomic plus molecular gas mass of $\sim 10^7$~M$_{\odot}$ (\citealt{Motte2003, NguyenLuong2011, Carlhoff2013}; \citetalias{Bihr2015}). Based on parallax measurements of water and methanol masers, the distance to W43 is $d=5.5\pm0.5$\,kpc \citep{Zhang2014}, placing it at 
the intersection of the Galactic bar with the first spiral arm \citep{NguyenLuong2011}.

\citetalias{Bihr2015} reported on extremely large  \ho column densities, $\sob 120-180 \ \mpc$,  in W43. 
As shown by \citetalias{Motte2014} and \citetalias{Bihr2015}, the observed \ho columns strongly exceed the columns theoretically expected from single cloud \tra transition models.
In such models (e.g., \citealt{Krumholz2009}; hereafter \citetalias{Krumholz2009}, \citealt{McKee2010}, and \citealt{Sternberg2014}; hereafter \citetalias{Sternberg2014}),
the \tra transitions are computed assuming a balance between far-UV photodissociation and molecular formation, and accounting for the rapid attenuation of the radiation field due to H$_2$ self-shielding and dust absorption.
For solar metallicity, the maximal predicted \ho columns are $\sim 10 \mpc$, far less than observed in W43.

In this paper we analyze the \ho columns in W43.
We employ the \citetalias{Sternberg2014} analytic model for the equilibrium \ho columns produced in the transition layers, as functions of the field intensity, gas density, grain properties and metallicity.
The \citetalias{Sternberg2014} and \citetalias{Krumholz2009} models are very similar in their predictions for the \ho columns (see  \S 4 in \citetalias{Sternberg2014} for a comparison of the planar versus spherical geometry). 
We show that the simplest and most likely explanation for the large observed \ho columns is 
% that the simplest and perhaps most probable explanation is 
 that several \tra transition layers are superimposed along the sightlines.
 This possible interpretation was already noted by \citetalias{Motte2014}.
 
The structure of the paper is as follows.
 We present the observations in \S \ref{sec: Observations}.
In \S \ref{sec: Analysis} we analyze the data and estimate the number of \tra transition layers.
We discuss the implications of our analysis and alternative scenarios in \S \ref{sec: discussion}.
We conclude in \S \ref{sec: conclusions}.
%model the \ho and \hh constituents of a gas cloud illuminated by far-UV radiation, assuming an H$_2$ formation-destruction steady-state.
%The far-UV photons lead to H$_2$ photodissociation and produces $\ho$.
%With increasing cloud depth, the radiation is getting absorbed by dust and by the H$_2$ molecules, until a sufficient attenuation is attained and an \tra transition occurs.
%Various models invoke different assumptions, such as the cloud geometry (e.g., slabs versus spheres), irradiation geometry (e.g.~isotropic versus beamed), density structure, etc.. 
%However, all steady-state \tra models agree that the \ho column density cannot strongly exceed $\sho \sim 10 \ \mpc$ (for solar metallicity). 
%This is a result of the far-UV exponential absorption by dust-grains (corresponding to a dust opacity $\tau \gtrsim 1$).
%}

%(\citetalias{Krumholz2009}, \citetalias{Sternberg2014})
%\citetalias{Bihr2015} proposed that perhaps the intense far-ultraviolet (UV) field expected in W43 is responsible for the ``too-large" \ho columns.
%\citetalias{Motte2014} argued that W43 might be experiencing dynamical mass accumulation via gas streams.
%In their picture, the large observed \ho columns are the result of the atomic accumulated gas, which did not yet have time to convert into H$_2$ (i.e., a non steady-state scenario).
\begin{figure*}
	\centering
	\includegraphics[width=1\textwidth]{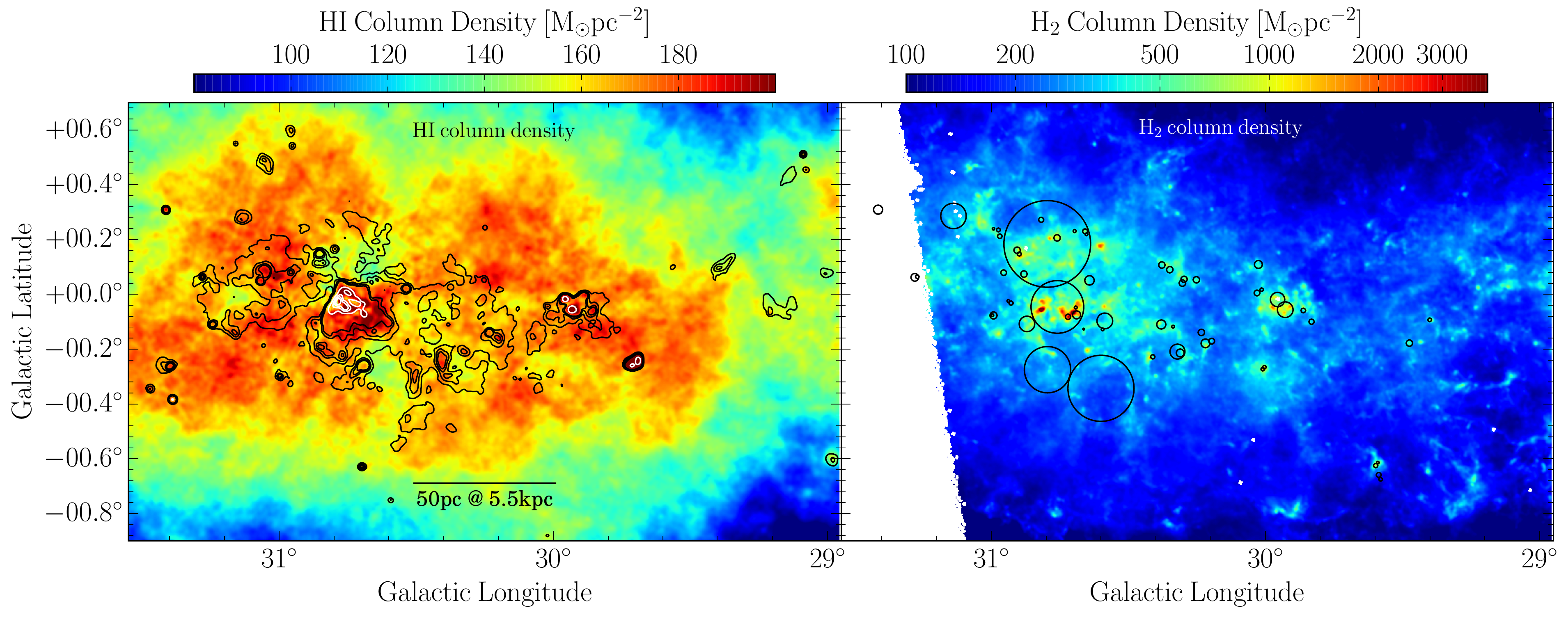}
	\caption{ 
Left panel: HI column density taken from \citet{Bihr2015}, corrected for optical depth effects and the continuum emission. The contours indicate the 1.4\,GHz continuum emission (black contours: 10, 30 and 70\,K, white contours: 200, 400, 600 and 800\,K). Right panel: \hh column density, based on Herschel dust observations from \citet{Nguyen2013}. The black circles indicate HII regions taken from \citet{Anderson2014}.
	}
	% this figure was generated using the script: 
	\label{fig:observations}
\end{figure*} 

\section{Observations}
\label{sec: Observations}
The analyzed \ho data is based on observations of ``the $\ho$, OH, Recombination line survey of the Milky Way" (THOR; \citetalias{Bihr2015}; \citealt{Beuther2016}). For the \ho column density of W43, it is crucial to correct for 21 cm optical depth, as well as for weak diffuse continuum emission, otherwise the measured HI mass is underestimated by at least a factor of 2.4 (\citetalias{Bihr2015}). 
The measured (corrected) \ho column density map is shown in the left panel of Fig.~\ref{fig:observations}. 
The \ho column density typically ranges between 120 and 180 $\mpc$.
We note, that the corrected \ho column density does not decrease toward the center as reported by previous studies \citep{NguyenLuong2011}, assuming optically thin \ho emission.
% The previous optically thin \ho maps supported a picture of a coherent \ho envelope-like structure, surrounding the central region of W43 (\citetalias{Motte2014}).
% Molecular hydrogen cannot be observed directly. 
The molecular gas, as inferred from dust observations is shown in the right panel of Fig.~\ref{fig:observations} \citep{Nguyen2013}. It is much clumpier than the $\ho$, and may be decomposed into $\sim 20$ sub-clouds \citep{Carlhoff2013}. The H$_2$ column densities typically range within $\sim 100-500 \ \mpc$ with peak values reaching $\sim 3000 \ \mpc$ at the cores (\citealt{Nguyen2013}, \citealt{Carlhoff2013}, \citetalias{Bihr2015}).

To appreciate the high level of star formation activity in the region, the white circles in the right panel of Fig.~\ref{fig:observations} indicate \hii regions, taken from the \citet{Anderson2014} catalog, having distance estimates consistent with W43 ($d\approx4-8$~kpc).  
The contours in the left panel indicate the 1.4\,GHz continuum emission.
The relatively high dust temperatures $T_{\rm dust} \approx 23-27$~K \citep[higher than typical $T_{\rm dust} \lesssim 20$~K;][]{Reach1995, Draine2011} reported by \citet[][]{Nguyen2013}, also suggest intense far-UV irradiation (see \S \ref{sub: aG in W43}).

\section{Analysis}
\label{sec: Analysis}

We analyze the observed \ho column densities in W43, using the \citetalias{Sternberg2014} steady-state \tra transition model.
The \citetalias{Sternberg2014} model assumes optically thick, uniform-density slabs, irradiated by beamed or uni-directional far-UV flux. 
For a two sided slab and isotropic irradiation, the \ho column density is 
\begin{equation}
\label{eq: Sig_HI}
\sho \ = \  \frac{6.71}{\st} \ \ln \left( \frac{\aG}{3.2} \ + \ 1 \right) \ \mpc \ ,
\end{equation}
where $\st \equiv \sg/(1.9 \times 10^{-21} \ {\rm cm^2})$ is the dust absorption cross section per hydrogen nucleus in the Lyman-Werner (LW) dissociation band (11.2-13.6 eV), normalized to a fiducial Galactic value (i.e., typically $\st \approx 1$), $\alpha$ is the ratio of the unshielded H$_2$ dissociation rate to H$_2$ formation rate, and $G$ is an average H$_2$-self shielding factor in dusty clouds (see \citetalias{Sternberg2014}, and \citealt{Bialy2016} for a thorough discussion).
%where $Z'$ is the dust-to-gas ratio in units of Galactic, and $\phi$ is a parameter encapsulating the properties of dust absorption, and is of order unity (refs). 
For a \citet{Draine1978} interstellar radiation spectrum shape, and H$_2$ formation on dust grains
\begin{equation}
\label{eq: aG param with R}
\aG \ = \ 2.0   \ I_{\rm UV} \left(\frac{30 \ {\rm cm^{-3}}}{n} \right) \left( \frac{9.9}{1+8.9 \st} \right)^{0.37}  \ .
\end{equation}
Here $n$ is the volume density and $I_{\rm UV}$ is the interstellar radiation intensity relative to the \citet{Draine1978} field.
For typical cold neutral medium (CNM) conditions, $\aG \simeq  2.6$, and is weakly dependent on $\st$ \citep{Bialy2016}.
Importantly, Eq.~(\ref{eq: Sig_HI}) is for a {\it single} two-sided slab (an $\ho$-$\hh$-$\ho$ ``sandwich", hereafter referred by ``\tra transition layer").

Given the large extent of W43 ($\sim 100 \times 200$~pc), the wealth of molecular structures 
  and the many embedded radiation sources (Fig.~\ref{fig:observations}, \citealt{Carlhoff2013}), we argue that the observed \ho gas is probably not attributed to a single \tra transition layer but is composed of several transition layers, superimposed along the W43 sightlines.
% This naturally explains the large \ho columns observed in W43.
The idea of several transition layers is further supported by the \ho spectrum which shows several distinct peaks (see Fig.~10 in \citetalias{Bihr2015}), and also by observations of [C {\small II}] 158$\mu$m emission \citep[][as further discussed in \S\ref{sec: discussion}]{Shibai1991}.
The \ho spectrum saturates at an optical depth of $\sim 3$, and deeper observations of W43 are expected to reveal more structures.
In \S \ref{sec: discussion}  we also discuss alternative scenarios for the large observed \ho columns in W43.

Assuming \Nh transition layers along a sightline, the observed \ho column density is given by
\begin{equation}
\label{eq: Sig_obs}
\sob \ =  \ \phi \ \Nh \ \sho \ ,
\end{equation}
where $\phi$ is a geometrical factor of order unity, and $\sho$ is the column density of a single \tra transition layer as given by Eq.~(\ref{eq: Sig_HI}).
Assuming that the \tra transition layers are slabs that are randomly oriented, $\phi=1$.
This is because a slab that is tilted by an angle $\theta$ relative to the plane of the sky, will have a column density $\propto 1/\cos(\theta)$, but will also have an area filling factor $\propto \cos(\theta)$.

In Fig.~\ref{fig: Sigma_aG} we plot $\sho$ as a function of $\aG$ as given by Eq.~(\ref{eq: Sig_HI}). 
Overplotted are the observed $\sob \approx 120-180 \ \mpc$  (horizontal strip).
The vertical strip is the realistic $\alpha G$ range estimated for W43 (see \S \ref{sub: aG in W43}).
Fig.~\ref{fig: Sigma_aG} (and Eq.~\ref{eq: Sig_HI}) shows that for a single cloud model to fit the observations, unrealistically large values of $\alpha G$ are required.
This remained an unresolved puzzle in \citetalias{Bihr2015}.
In the rest of this section we estimate  $\aG$ and obtain the mean \ho column for a single transition layer in W43.
We then constrain the number of transition layers along the sightlines using the observed \ho columns combined with Eq.~(\ref{eq: Sig_obs}).

 \subsection{$\aG$ in W43}
  \label{sub: aG in W43}
 To constrain $\aG$ in W43, we should approximate the UV intensity $I_{\rm UV}$ and the volume density $n$ for the \ho gas of W43 (see Eq.~\ref{eq: aG param with R}).
%The high level of star formation activity and the relatively warm dust suggest an enhanced far-UV radiation intensity in W43, relative to the standard Galactic value $I_{\rm UV}=1$.
Adopting $T_{\rm dust} = 23-27$~K (\S \ref{sec: Observations}), and assuming thermal equilibrium for the dust grains, $I_{\rm UV}=(T_{\rm dust}/T_0)^6$ with $T_0=17.5$~K \citep{Draine2011}
%\footnote{The normalization factor appropriate for silicate grains of radii $a=0.1$~$\mu$m (the value weakly depending on radius), was modified to correct for the different normalization of the \citet{Draine1978} versus the \citet{Mathis1983} interstellar radiation fields.}
, we obtain $I_{\rm UV}=5.2-13.5$.
The modeling of the dust temperature is uncertain to within a factor of $\sim 2$, therefore we consider a wider range for $I_{\rm UV}$, and adopt $I_{\rm UV}=3-30$.
% XXX higher LW flux (up to 10 compared to mean ISRF) for an OB BB spectrum.
% Accounting for UV attenuation \citep{Glover2012, Beuther2014} deduced somewhat higher $I_{\rm UV}$, up to $\sim 100$.
% For our purposes we consider the conservative range $I_{\rm UV}=10-100$ for W43.

%For example, the rich OB and Wolf-Rayet stars cluster present within W43-main, has a luminosity of $L \approx 3.5 \times 10^{6} \ L_{\odot}$ (refs) and produces a flux of LW radiation $F_{\rm LW} = 1.5 r_{\rm pc}^{-2}$~erg~cm$^{-2}$~s$^{-1}$, where $r_{\rm pc}$ is the radial distance in parsecs. 
%This radiation is equivalent to an isotropic ISRF of $I_{\rm UV}=150$ at a distance of 10~pc, or $I_{\rm UV}=18$  at 30 pc. Various HII regions identified in W43 are shown in Fig.~XXX  (refs).

We shall now obtain an approximation of the volume density $n$. 
Theoretically, the cold neutral medium (CNM;  $n \approx 30$~cm$^{-3}$; \citealt{Draine2011}), is expected to dominate the shielding and the \tra transition (\citetalias{Krumholz2009}, although see \citealt{Bialy2015}).
For a CNM gas at thermal equilibrium with warm neutral medium (WNM), 
the volume density of the CNM is an increasing function of $I_{\rm UV}$ \citep{Wolfire2003}.
For $I_{\rm UV}=3-30$, \citet{Wolfire2003} suggest $n \sim 30 -100$~cm$^{-3}$.
% \citealt{Bialy2015} found that the \ho envelopes in the Perseus molecular cloud might be smaller by a factor of 2-10 than the CNM value for Perseus.
From an observational point of view, by averaging the observed \ho column densities over elliptical annuli and dividing by the equivalent radius, \citetalias{Bihr2015} estimated $n \approx 10-20$~cm$^{-3}$ for the \ho gas in W43.
This procedure assumes a uniform density gas. 
For a clumpy medium the \ho density values will increase.
In summary, given the uncertainties, we adopt $n=10-100$~cm$^{-3}$ for the atomic gas of W43.

\begin{figure}[t]
	\centering
	\includegraphics[width=0.5\textwidth]{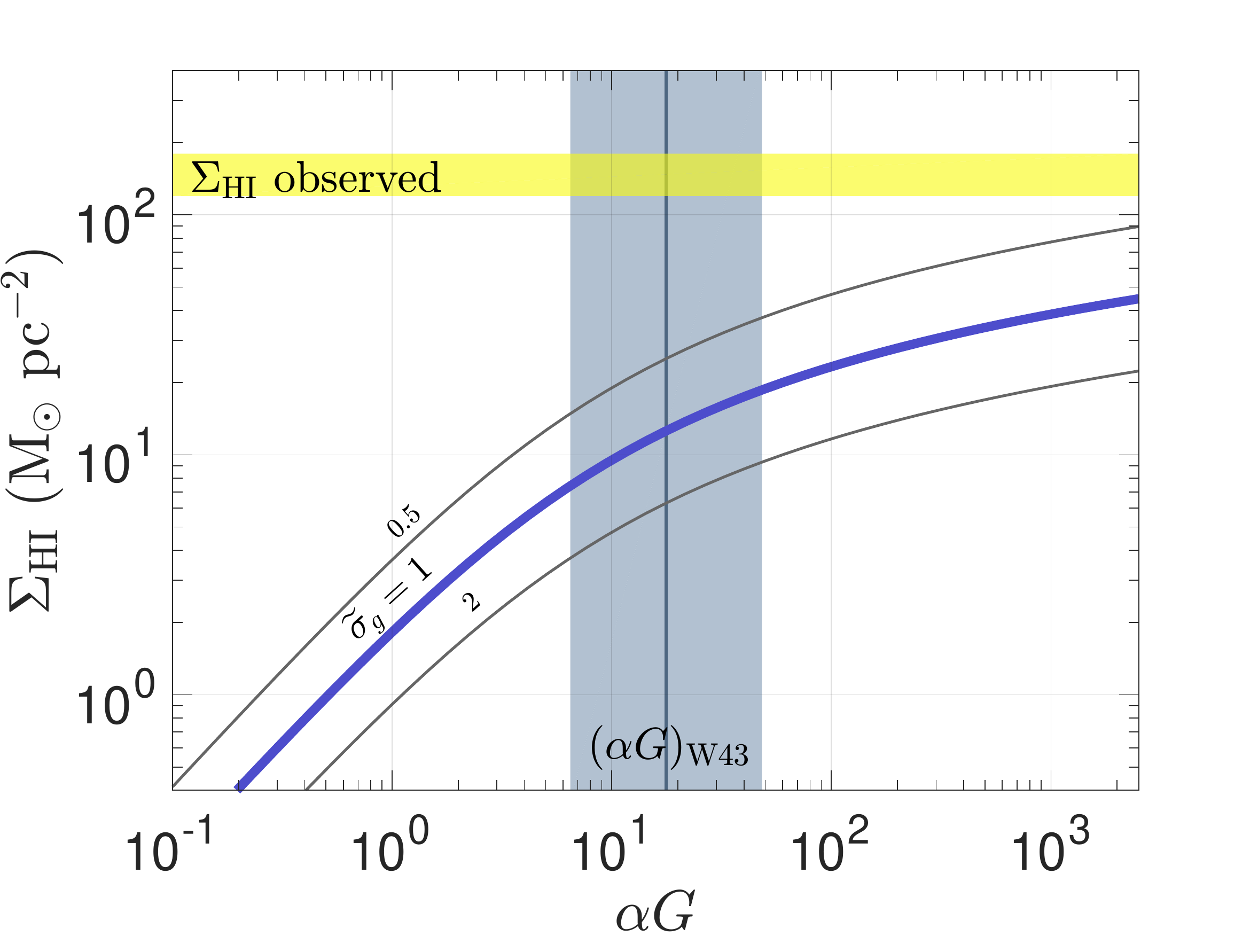}
	\caption{The \ho column density of a single \tra 
    transition layer as a function of the $\aG$ parameter, as given by Eq.~(\ref{eq: Sig_HI}) assuming $\st=1$ (wide blue curve) and factor of two variations about $\st=1$ (thin gray).
	The vertical line and shaded strip are the median $\aG$ and the 68.3\% range (i.e.~``1-sigma" error) for W43.
    The observed \sob in W43 (yellow shading) are larger by factors of $\sim 10-20$ than the theoretical curve.
	}
	% this figure was generated using the script: 
	\label{fig: Sigma_aG}
\end{figure}

We use Eq.~(\ref{eq: aG param with R}) and evaluate the $\aG$ probability distribution function (PDF) assuming that $\log n$ and $\log I_{\rm UV}$ are uniformly distributed within the adopted ranges, giving
\begin{equation}
\label{eq: aG_W43}
(\alpha G)_{\rm W43} \ = \ 17.7^{+30.8}_{-11.2}
\end{equation} for the median $\aG$ and the ``1-sigma" error (see Appendix).
The median and 1-sigma range are shown by the vertical line and shading in Fig.~\ref{fig: Sigma_aG}. 
% $\aG=25.1$ value, and the $1, 2, 3$-sigma ranges, $\aG=6.5-48.4$, $2.9-109.0$, and $2.0-160.0$, enclosing the 68.3, 95.5, and 99.7 percentiles about the median, respectively.
This $\aG$ was evaluated at $\st=1$.
 Although the value of $\aG$ depends on the assumed $\st$, the dependence is weak
% ($\aG \propto \st^{-0.37}$ for $\st>0.1$ and is independent of $\st$ for $\st<0.1$; Eq.~\ref{eq: aG param with R}) 
so for the case of simplicity we show the $\aG$ ranges only for $\st=1$.
%The adopted value of $\st$ has a much stronger effect on the \ho column through the inverse dependence in Eq.~\ref{eq: Sig_HI}, as we discuss below.

 \subsection{The number of \tra transition layers}
  \label{sub: no of transitions}
%The physical conditions may change with location (e.g., stronger radiation near the OB clusters, varying volume density etc.)
%- these variations are contained in our uncertainty ranges of $\aG$ (\S \ref{sub: aG in W43}).
Adopting the $\aG$ range given by Eq.~(\ref{eq: aG_W43}),
the column density of a single \tra transition layer is $\sho = 13^{+6}_{-5}/\st \ \mpc$ (Eq.~\ref{eq: Sig_HI}).
Following Eq.~(\ref{eq: Sig_obs}), the number of \tra transition layers is 
\begin{equation}
\label{eq: Nh}
\Nh \ =  \ 13^{+9}_{-4} \ \left( \frac{\sob}{160 \ \mpc} \right) \ \st
\end{equation}
where \sob is an observed \ho column.
%, and $\st$ is the assumed dust absorption cross-section (recall, $\st=\sg/(1.9\times 10^{-21}$~s$^{-1})$ and is $\approx 1$ for standard conditions). 
%For the observed $\sho=120$ to $180 \mpc$, and assuming $\st=1$, $\Nh=6-16$ and $\Nh=9-23$ respectively.
In Eq.~(\ref{eq: Nh}), the typical dust absorption cross section is $\st=1$ (corresponding to $\sg=1.9\times 10^{-21}$~cm$^2$).
The value of $\st$ depends on the dust-to-gas ratio and the composition and size distribution of the dust grains. 
The metallicity and the dust-to-gas ratio typically increase with decreasing galactocentric distance, suggesting that $\st$ is possibly larger by up to a factor of $\sim 2$ in W43. 
The value of $\st$ depends also on the shape of the extinction curve. 
The fiducial value assumes $R_V \equiv A_V/E(B-V)=3.1$ (appropriate for diffuse gas $n\sim 100$~cm$^{-2}$), in dense regions ($n \gtrsim 10^3$~cm$^{-3}$) the dust-extinction curve becomes less steep towards the UV, and $\st$ may decrease by up to a factor of $ \approx 2$ \citep{Draine2003}.
Assuming a $\st=0.5-2$ uncertainty range and reevaluating the PDF of $\aG$, $\sho$ and $\Nh$, we obtain a larger median value and uncertainty range, $\Nh=13^{+13}_{-6}(\sob/160 \ \mpc)$.

Based on this analysis, a typical column density fluctuation in the \ho map may be approximated by $\Delta \sob/\sob \sim 1/\Nh \approx 5-12 \%$, arising from  sightlines that differ by a single transition layer. These values are in agreement with the observed \ho map shown in Fig.~\ref{fig:observations}.

The length-scale of W43 along the line-of-sight may be approximated, through $\Delta z = \lambda_{\ho} \Nh$, where $\lambda_{\ho}$ is the characteristic \ho scale, given by
\begin{equation}
\label{eq: HI-scale}
\lambda_{\ho} \ = \ 2 \ \frac{1}{\sg n} \ = \ 11.4 \ \frac{1}{\st} \left(\frac{\rm 30 \ cm^{-3}}{n}  \right) \ {\rm pc} \ ,
\end{equation} where the factor of two is for the two sides of the slab.
For $\Nh = 9-21$ and $n=30$~cm$^{-3}$, $\Delta z \sim 100-250$~pc, comparable to the projected dimensions of W43.
This is only a lower limit on $\Delta z$ since the \tra transition layers may be spatially separated with diffuse atomic and ionized gas.

\section{Discussion}
\label{sec: discussion}

We have shown that the observed \ho column densities in W43 are naturally explained by a model of $\sim 9-21$ \tra transition layers superimposed along the W43 sightlines.
Several transition layers are also expected from the clumpy molecular structure revealed by CO and dust, the high star formation activity and wealth of embedded radiation sources.

Several transition layers along the W43 sightlines were also inferred from [C {\small II}] 158$\mu$m observations \citep{Shibai1991}.
Assuming thermal balance between [C {\small II}] 158$\mu$m cooling and photoelectric heating, \citep{Shibai1991} found $\Nh \sim 5$ (see their \S 4.4).
However, they assumed a different geometry for the clouds (thin shells).
To compare with our model, we modify their formula (their third Eq.~in \S 4.4) to match our randomly-oriented-slab geometry\footnote{We divide their prefactor by 2. An additional factor of 1.7 comes from the normalization of the \citet{Habing1968} radiation fields compared to the \citet{Draine1978} field.}, giving 
\begin{equation}
I_{[CII]} \ \simeq 5.1 \times 10^{-6} \ \Nh \ I_{\rm UV} \ {\rm erg \ cm^{-2} \ s^{-1} \ sr^{-1}} \ .
\end{equation} 
\citet{Shibai1991} reported observations of $I_{[CII]}=7 \times 10^{-4}$~erg~cm$^{-2}$~s$^{-1}$~sr$^{-1}$ implying $\Nh=13.7(I_{\rm UV}/10)^{-1}$, consistent with our result $\Nh=13^{+9}_{-4}$.

\subsection{Model and Observational Limitations}
\label{sub: limitations}
The \ho map of \citetalias{Bihr2015} was obtained by integrating 
the 21 cm observations over the ``complete" velocity range of W43, $v_{lsr}=60-120$~km~s$^{-1}$ \citep{NguyenLuong2011}.
If part of the observed \ho gas is unrelated to W43, or is very diffuse and extended, 
then \sob should be decreased, and the value of \Nh will decrease accordingly.
However, even if we consider only half of the observed $\sho$, still several \tra transitions are required.

In contrast to this, \sob could also be higher: \citetalias{Bihr2015} used the optical depth $\tau$ to correct $\sob$. As the used absorption spectrum saturates, the measured optical depth is a lower limit and hence \sob is a lower limit as well.

The \citetalias{Sternberg2014} theory assumes chemical equilibrium.
The longest time-scale involved is  
% For well shielded gas (that will eventually become molecular), the time-scale is
%\begin{equation}
%\label{eq: HI-time}
%\tau_{\rm eq} \ = \ \frac{1}{D+2Rn} \ ,
%\end{equation} where $D$ is the local (attenuated) H$_2$ photodissociation rate.
%In free space (no attenuation) $D=D_0 \gg 2Rn$, and $\tau_{\rm eq}$ is then the photodissociation time-scale.
%For \citet{Draine1978} ISRF $D_0=5.8 \times 10^{-11} \ I_{\rm UV}$~s$^{-1}$ (\citetalias{Sternberg2014}), 
%$\tau_{\rm eq} =  540/I_{\rm UV}$~yr, and the system approaches steady-state swiftly.
%In highly shielded regions $D \ll 2Rn$ and the time-scale is 
the H$_2$ formation-time, $\tau_{\rm H_2}=1/(2Rn)$, where $R$ is the H$_2$-formation rate-coefficient.
For $R = 3 \times 10^{-17}$~cm$^{-3}$~s$^{-1}$ and $n=30$~cm$^{-3}$, $\tau_{\rm H_2}=18$~Myr.
In a turbulent medium the H$_2$ formation time may be much shorter, $\sim 1-2$~Myrs \citep{Glover2007}.
% For weakly shielded gas (that will eventually become atomic) the time-scale is the photodissociation time and is very short. 
% For a free-space photodissociation rate of $4.9 \times 10^{-11}$~s$^{-1}$ (\citetalias{Sternberg2014}), $\tau_{\rm H_2}=650$~yr.
%Enhanced H$_2$ formation rates on surfaces of PAHs may also decrease the H$_2$ formation time by a factor of $\sim 3$ \citep{Habart2003}.

The \citetalias{Sternberg2014} model assumes a uniform-density gas for the \ho shielding envelopes.
In real astrophysical environments the gas is highly turbulent, producing large density fluctuations.
However, Bialy, Burkhart \& Sternberg (in prep) find that even in highly turbulent supersonic gas, the median \sho value remains very close to the homogeneous solution given by Eq.~(\ref{eq: Sig_HI}). 
Moreover, for large $\aG$ the spread in \sho is small (typically within a factor of $\sim$ 2).

Eq.~(\ref{eq: aG param with R}) for $\aG$ assumes an H$_2$ formation rate coefficient of $R=3\times 10^{-17}$~cm$^{3}$~s$^{-1}$ and a \citet{Draine1978} spectrum shape. The value of $R$ is highly uncertain.
However, the fact that the dependence of $\sho$ on $\aG$ is logarithmic (Eq.~\ref{eq: Sig_HI}), makes our analysis robust to variations in all parameters entering $\aG$; the UV intensity, spectrum shape, volume density, H$_2$ formation rate coefficient and the H$_2$-self shielding function.
%The dependence on $\st$ is on the other hand strong (linear), and the typical factor of $\sim 2$ uncertainty in $\st$ (\S \ref{sub: no of transitions}) introduces a significant uncertainty to the value of $\Nh$ (Eq.~\ref{eq: Nh}).

\subsection{Alternative explanations for the large \ho columns}
\label{sub: alternative scenarios}
\citetalias{Motte2014} and \citetalias{Bihr2015} analyzed the observed \tra column densities using the \citetalias{Krumholz2009} equilibrium model and also found that the observed \sho values are far in excess to those predicted theoretically.
This is not surprising since like the \citetalias{Sternberg2014} model, the \citetalias{Krumholz2009} model is also a steady-state model that is applicable to a single cloud.
The \citetalias{Motte2014} and the \citetalias{Bihr2015} studies 
proposed alternative explanations for the theory-observations discrepancy.

\citetalias{Bihr2015} suggested that the intense UV radiation field in W43 may account for the large observed $\sob$.
Following Eq.~(\ref{eq: Sig_HI}) we see that for a single \tra transition to reproduce the observed \ho column densities, $\aG$ must be extremely large. For example, for $\sob=120-180 \ \mpc$, and assuming $\st=1$, $\aG=1.9\times10^8 - 1.4\times 10^{12}$ respectively, requiring unrealistically high $I_{\rm UV}$ to density ratios (see Eq.~\ref{eq: aG param with R}).
% The fact that the dependence on $\aG$ is logarithmic, makes our analysis robust to variations in the UV intensity, spectrum shape, volume density, H$_2$ formation rate coefficient and H$_2$-self shielding function (all the parameters  entering $\aG$; \citetalias{Sternberg2014}, \citetalias{Bialy2016}).

A single equilibrium transition layer may be consistent with observations if the dust absorption cross section ($\st$) is significantly reduced.
 For the 1-sigma $\aG$ range, $\st=0.05-0.11$ reproduce the observed $\sho=160 \ \mpc$.
 Such low dust absorption cross-sections may be a result of a reduced dust-to-gas ratio  or an abnormal population of dust grains which is selectively inefficient in far-UV absorption.
 However, this is unlikely given the very large deviation from typical ISM values.
 The effectiveness of dust absorption may also be reduced by dust-gas separation
 resulting from non-isotropic radiation \citep{Weingartner2001}.
 However, the time-scales involved are very long. For 0.1 $\mu$m grains $t=100 \ r_{pc}$~Myr, where $r_{pc}$ is the separation length in pc.

% Assuming a standard dust-to-gas ratio, \citet{Carlhoff2013} found a good agreement between the molecular columns obtained from $^13$CO observations and from Herschel dust observations. 
\citetalias{Motte2014} identified three velocity gradients in the position-velocity map of W43, possibly indicative of inflowing gas streams.
The large \ho columns were then interpreted by \citetalias{Motte2014} as dynamically accumulated atomic gas which has  not had time yet to convert into H$_2$.
In this scenario the accretion time-scale must be short compared to the H$_2$ formation time, $\tau_{\rm H_2}=1-20$~Myr  (\S\ref{sub: limitations}), thus requiring a minimum inflow rate of $\dot{M}_{\rm \ho}=2.7$-$0.15 \ {\rm M_{\odot} yr^{-1}}$ (assuming $M_{\ho}=2.7\times10^6 \ M_{\odot}$; \citetalias{Bihr2015}).
However, \citetalias{Motte2014} were unable to estimate the gas inflow rates.
Even if gas streams are present, they may be dense enough to already contain the equilibrium $\ho$/\hh interfaces, and contribute to the total number $\Nh \approx 13$, estimated in \S \ref{sub: no of transitions}.
%In fact, the streams reported by \citetalias{Motte2014} have significant molecular contributions.}

The alternatives of photodissociated multiple layers versus a non-equilibrium atomic inflow may be distinguished by the presence or absence of infrared (IR) H$_2$ line emission. 
%In our scenario, the \ho column density is a product of H$_2$ photodissociation by the Lyman-Werner (LW) photon flux (as opposed to, e.g., the time-dependent scenario discussed in \S \ref{sub: alternative scenarios}).
In the layers, far-UV radiation not only photodissociates H$_2$, but also populates excited H$_2$ rotational-vibrational levels, resulting in energetic cascade and IR emission.
In contrast, for an atomic inflow there are no molecules to excite.
Thus, our scenario predicts high IR-line emissions, with an integrated flux being proportional to the observed \ho column densities \citep{Black1987, Sternberg1988}.
Such observations could confirm the predictions of our multi-slab model.
To our knowledge, there are no current IR spectra of the H$_2$ ro-vibrational transitions in W43.

\subsection{Comparison to other \ho observations}
A much less extreme example of enhanced \ho columns may be found in the Perseus molecular cloud.
\citet{Lee2015} reported $\sob \approx 6-9 \ \mpc$, and \citet{Bialy2015} showed that these values are 2 to 3 times larger than what is theoretically expected for CNM conditions.
\citet{Bialy2015} showed that this discrepancy is alleviated if the \ho gas density in Perseus is lower than typical CNM, with $n \approx 2-10$~cm$^{-3}$.
The analysis of \citet{Bialy2015} assumed a single \tra transition layer.
Following the above discussion, a potentially alternative explanation for the ``too-large" \ho columns in Perseus would be that there are typically two to three \tra transition layers along the Perseus sightlines.
%However, this interpretation could be problematic because then we would expect to see large fluctuations in the observed \ho column density map, of $\Delat \sho/\sho \approx 1/\Nh = 30-50 \%$, inconsistent with the relatively smooth \ho map
%resulting from sightlines differing in the number of transition layers they probe.
%For example, for two sightlines probing one versus two transition layers, we would get a $\sim 50 \%$ difference in $\sho$, whereas the observed \ho map in Perseus is relatively uniform.
%On the contrary, in W43 we infer $\Nh = 9-21$, thus typical fluctuations expected in $\sho$ are of order $\sim 5-10 \%$ -  in agreement with the observed \ho map (Fig.~\ref{fig:observations}).

Extragalactic observations often find an \ho threshold of $\approx 10 \ \mpc$ \citep[e.g., ][]{Bigiel2008,Schruba2011}.
This threshold was explained in the framework of a steady-state \tra transition model (\citetalias{Krumholz2009}).
% above which the gas is fully molecular 
% \citet{Krumholz2009} presented a theoretical model for the \tra transition occurring in a spherical gas cloud, and found good agreement between the extragalactic observations and the theoretical model.
However, while the model applies to a single cloud,
the extragalactic observations, having resolutions of  $\sim$ kpc, are not able to resolve molecular clouds, and even not large molecular complexes such as W43.
Interpreting the \ho threshold as an \tra transition is therefore problematic. 
% The observed \ho columns cannot be associated with a single \tra transition.
The \ho threshold might be a result of two effects acting in opposite directions, (i) several \ho layers along the line of sight that increase the column density, and (ii) beam dilution by diffuse intercloud gas and WNM, that lowers the observed column densities \citep{Parmentier2016}.
% In this picture, the observed \ho threshold may be used to estimate the area filling factor of dense molecular cloud complexes.
% For example, assuming a negligible column density for the diffuse gas, and a column density $\sho_{,\rm comp}$  for a dense molecular complex, then the area filling factor of the complex is $f_A \leq \Sigma_{\rm thresh}/\sho_{,\rm comp}$. 
% For example, for a W43-like complex with  $\sho_{\rm comp} \sim 160$,  $f_A<0.06$.
% A more sophisticated treatment is required to draw further conclusions, and will be presented elsewhere.

\section{Conclusions}
\label{sec: conclusions}
The main results of our paper are as follows. 
\begin{enumerate}
\item The observed \ho column densities in W43, $\sob=120-180$~\mpc are very large compared to those predicted by equilibrium \tra theories, $\sho \sim 10 \ \mpc$ (\citetalias{Krumholz2009}; \citetalias{Sternberg2014}).
\item Realistic variations of the far-UV flux, spectrum shape, volume density, or the H$_2$ formation rate coefficient, cannot alleviate the theory-observation discrepancy.
\item The large observed \sob are naturally explained by our multi-slab model, in which $9-21$ transition layers are superimposed along the W43 sightlines. CO, dust and [C {\small II}] observations also support this model \citep{Shibai1991}.
%\item Some of the observed \ho columns may result from diffuse intercloud gas and fore- and back-ground contamination, lowering $\Nh$. However, still several \ho components are required. 
%\item A higher (lower) dust absorption efficiency increases (decreases) the number of \tra transition layers.
%\item $\Nh$ is proportional to the assumed value of the dust absorption cross-section $\st$.
%Typically $\st \approx 1$ with realistic variations of a factor of $\sim 2$.
%\item The time-dependent scenario and the low dust-absorption scenario are less likely to explain the large observed \ho columns in W43.
\item Time-dependent accumulation of \ho is not necessary to account for the large \ho columns observed. We predict the presence of IR H$_2$ line emissions from the multiple transition layers, in our picture.
Such emissions would be absent for accumulating inflowing atomic gas.
\item The \ho threshold often observed in extragalactic observations \citep[e.g.][]{Bigiel2008} may be a result of telescope beam averaging of (i) large column densities due to many \ho clouds along the sightlines, with (ii) additional diffuse low column density gas.
\end{enumerate}

\acknowledgments
We thank Jouni Kainulainen, Bruce Draine, Simon Glover and Sahar Shahaf for helpful suggestions and fruitful discussions. We thank the referee for helpful comments on our manuscript.
SB thanks the MPIA for visitor support where this research was carried out.
This work was supported in part by the PBC Israel Science Foundation I-CORE Program grant 1829/12.

\appendix
To obtain the PDF of $\aG$, and in particular the median $\aG=17.7^{+30}_{-11.5}$ (Eq.~\ref{eq: aG_W43}), where the error corresponds to the 68.3 percentile about the median (i.e., ``1-sigma"), we assume that $\log I_{\rm UV}$ and $\log (n/{\rm cm^{-3}})$ are  uniformly distributed within our adopted $I_{\rm UV}=3-30$ and $n=10-100$~cm$^{-3}$ ranges (\S \ref{sub: aG in W43}).
We have chosen a uniform distribution since it is the maximum entropy distribution.
To obtain the PDF of $\aG$, let us introduce the following random variables,  
\begin{align}
X &\equiv \log_{10} \left( \frac{I_{\rm UV}}{3} \right) \\
Y &\equiv -\log_{10} \left( \frac{n}{100 \ {\rm cm^{-3}}} \right) \\ 
Z &\equiv \log_{10} \left( \frac{\alpha G}{1.8} \right) \ .
\end{align} 
With these definitions, $X$ and $Y$ are uniformly distributed within the $[0,1]$ range, and $Z=X+Y$ (assuming Eq.~\ref{eq: aG param with R} with $\st=1$).
The PDF of $Z$ is thus a convolution of the $X$ and $Y$ distribution, and results in the symmetric triangular distribution, with a mean of 1 and a standard deviation $1/\sqrt{6}$.
This gives $\aG=17.7^{+30}_{-11.5}$ for the median and 68.3 percentile about the median. 
The 95.5, and 99.7 percentiles correspond to $\aG$=2.9-108.4, and 2.0-157.0, respectively 
(the ranges are symmetric about the median in log space).
The PDF of $\aG$ may also be obtained numerically by generating large amounts of random pairs for \{$\log I_{\rm UV}$,$\log (n/{\rm cm^{-3}})$\}, and then using Eq.~(\ref{eq: aG param with R}) to obtain the $\aG$ PDF. 
We have followed such a numerical procedure to 
verify the above analytic result.

\end{document}